\documentclass[conference]{IEEEtran}
\IEEEoverridecommandlockouts
\usepackage{cite}
\usepackage{amsmath,amssymb,amsfonts}
\usepackage{algorithmic}
\usepackage{graphicx}
\graphicspath{{images/}}
\DeclareGraphicsExtensions{.png}
\usepackage{textcomp}
\usepackage{xcolor}
\usepackage{subfig}
\usepackage{multirow}

\usepackage{tikz}
\usepackage{textcomp}
\usepackage{hyperref}
\usepackage{lipsum}

\def\BibTeX{{\rm B\kern-.05em{\sc i\kern-.025em b}\kern-.08em
    T\kern-.1667em\lower.7ex\hbox{E}\kern-.125emX}}

\newcommand\copyrighttext{%
  \footnotesize \textcopyright 2019 IEEE. Personal use of this material is permitted. Permission from IEEE must be obtained for all other uses, in any current or future media, including reprinting/republishing this material for advertising or promotional purposes, creating new collective works, for resale or redistribution to servers or lists, or reuse of any copyrighted component of this work in other works.
Accepted paper for 2019 8th International Conference on Affective Computing and Intelligent Interaction (ACII). For citation and published version DOI, see below:\\
J. O’Dwyer, N. Murray, and R. Flynn, “Eye-based Continuous Affect Prediction,” in 2019 8th International Conference on Affective Computing and Intelligent Interaction (ACII), IEEE, Sep. 2019, pp. 137–143, isbn: 978-1-7281-3888-6. doi: \href{https://doi.org/10.1109/ACII.2019.8925470}{10.1109/ACII.2019.8925470}
}
\newcommand\copyrightnotice{%
\begin{tikzpicture}[remember picture,overlay]
\node[anchor=south,yshift=10pt] at (current page.south) {\fbox{\parbox{\dimexpr\textwidth-\fboxsep-\fboxrule\relax}{\copyrighttext}}};
\end{tikzpicture}%
}

\begin{document}

\title{Eye-based Continuous Affect Prediction
\thanks{This work was supported by the Irish Research Council (Grant No. GOIPG/2018/2030).}
}
\author{\IEEEauthorblockN{Jonny O'Dwyer, Niall Murray, Ronan Flynn}
\IEEEauthorblockA{\textit{Deptartment of Computer \& Software Engineering} \\
\textit{Athlone Institute of Technology}\\
Athlone, Ireland \\
j.odwyer@research.ait.ie, nmurray@research.ait.ie, rflynn@ait.ie}
}

\maketitle
\copyrightnotice

\begin{abstract}
Eye-based information channels include the pupils, gaze, saccades, fixational movements, and numerous forms of eye opening and closure. Pupil size variation indicates cognitive load and emotion, while a person's gaze direction is said to be congruent with the motivation to approach or avoid stimuli. The eyelids are involved in facial expressions that can encode basic emotions. Additionally, eye-based cues can have implications for human annotators of emotions or feelings. Despite these facts, the use of eye-based cues in affective computing is in its infancy, however, and this work is intended to start to address this. Eye-based feature sets, incorporating data from all of the aforementioned information channels, that can be estimated from video are proposed. Feature set refinement is provided by way of continuous arousal and valence learning and prediction experiments on the RECOLA validation set. The eye-based features are then combined with a speech feature set to provide confirmation of their usefulness and assess affect prediction performance compared with group-of-humans-level performance on the RECOLA test set. The core contribution of this paper, a refined eye-based feature set, is shown to provide benefits for affect prediction. It is hoped that this work stimulates further research into eye-based affective computing.
\end{abstract}


\section{Introduction}
A connection between the eyes and displays of emotion has been accepted for many years \cite{b1}. Eye-based cues are of increasing interest to the research community for automatic emotion classification and affect prediction. Such cues include eye gaze, eye saccades, pupillometry, fixational eye movements, and various forms of eye opening and closure events. The eyes can provide information on attention \cite{b2, b3, b4}, perception \cite{b5, b6}, social and emotional cues \cite{b5, b6, b7, b8, b9, b10}, cognitive load \cite{b11, b12}, locomotion \cite{b2}, and mental and physical pathology \cite{b3, b9, b10, b13}.

While the pupil size is known to vary under environmental, pathological and pharmacological conditions \cite{b13}, there is a body of evidence suggesting its efficacy for studies of neuropsychologic and affective responses in healthy inidviduals. Pupillometry studies from psychology showed the pupils to be responsive during emotional arousal \cite{b5, b6} and monetary incentive or penalty (therefore exhibiting valence potential) \cite{b14} during a memory task. In a neurophysiological study \cite{b10}, it was demonstrated that the pupils were responsive to autonomic nervous system stimulation, which is known to generate response output under numerous emotional states \cite{b15}. Neuropsychological evidence has been provided for pupilliary responses to reward expectation (positive valence event) \cite{b9}.

Eye gaze, the line of sight between an individual and an object of fixation, has been referred to as central to social understanding \cite{b3}. The shared signal hypothesis \cite{b7, b8} postulates that one's gaze is congruent with their emotional display if the gaze signal matches their underlying motivation to approach or avoid stimuli, direct gaze anger being a congruent threat cue for example. Eye-based cues, in addition to allowing subjects outwardly display their attention and/or social signal, can also have effects on social and emotional decoders or perceivers. Hess \cite{b16} reported behavioural changes in subjects who viewed (gazed at) image stimuli that had different pupil sizes; images with larger pupils were perceived to be more attractive than those with smaller pupils. Direct gaze (subject gazing directly at you) observation has been shown to contribute to attentional blink (AB) \cite{b17}. AB is where a subject's attention at a later duration in time is reduced or degraded due to stimuli received at an earlier point in time. In affective computing, models are often trained using audio-visual input features (interacting subject output). Additionally, ground-truth target values are generated by human annotators who look at and listen to (receive input from) audio-visual output provided. Therefore, it is important to understand eye-based cues from the perspective of both emotion encoders (interacting subjects) and emotion decoders (ground-truth annotators).

Blinks, winks, complete eye closure and partial eye opening or closure events are involved in certain eye gazes and saccades \cite{b18}, and facial expressions of emotion \cite{b19}. Additionally, it was suggested that eye gaze and blink share common signalling pathways \cite{b4}, which makes incorporation of eye closure and blink events important for a complete investigation of potential eye-based cues for affective computing. Eye-based cue estimation from video frames is now possible due to advances in computer vision models and tools \cite{b20}.

Based on identified trends and research opportunities in the literature, this work presents a comprehensive study of eye-based cues for affective computing, in particular for arousal and valence prediction. The core contribution of this paper is a proposal for feature sets from an alterative modality that make use of the aforementioned eye-based information channels for the purpose of continuous affect prediction. A secondary contribution of this paper includes the investigation of a feature selection method that incorporates annotator delay into the selection process. Both contributions are validated by way of continuous arousal and valence prediction experiments on the RECOLA \cite{b21} corpus. A practical use analysis combining the proposed eye-based feature sets with speech is also carried out. Code and feature sets used in the experiments will be made available to the research community on GitHub.

The remainder of this paper is structured as follows. Related work is presented in Section II. Data and tools for the experiments are detailed in Section III. In Section IV, the methods for feature selection and evaluation are described. Experimental results, along with discussion, are provided in Section V and the paper is concluded in Section VI.

\section{Related Work}
In affective computing, eye-based features are being investigated for emotion classification, stress detection, dyadic negotiation, dimensional affect prediction and psychopathology recognition. Emotion classification from facial expression images was carried out both in the presence and absence of direct/averted gaze features in \cite{b23}. Classification improvements for angry (5\%), sad (7.5\%), fear (5\%) and disgust (2.5\%) were observed relative to the facial expression-based system without the direct/averted gaze features. The presence of eye blink features was used for frustration classification in \cite{b24}, achieving 79.17\% accuracy for leave-one-subject-out validation. Positive/negative/neutral emotion classification was implemented using eye gaze and pupil features in \cite{b25}. An EyeLink 1000 eye tracking device was used in \cite{b25} to gather measurements from individuals observing image stimuli from \cite{b26} and a decision tree neural network was able to classify the individual's responses correctly at a 53.6\% rate on a subject-independent basis. Respondent reactions to negotiation offers were successfully predicted using a multimodal system including speech, eye gaze, head pose and smile features at a rate of 70.8\% average for 3x4-fold cross-validation in \cite{b27}. For unimodal arousal and valence classification using SVM, Soleymani et al. \cite{b22} used eye-based features, including statistics and spectral power calculations from the descriptors: pupil diameter, gaze distance, eye blinking, and $x$ and $y$ gaze coordinates gathered using a Tobii X120 eye tracking device. In their work, the eye-based features performed best when compared to electroencephalogram (EEG) and peripheral physiology measures. From the results in \cite{b22} the bimodal fusion of EEG and eye-based features performed best overall (arousal = 67.7\%, valence = 76.1\%). Eye gaze was combined with speech in \cite{b28}, where a feature set similar to that of \cite{b22} was used. Additional statistics were gathered for eye scan paths and eye closure features were measured by frame counts instead of time. Results achieved in \cite{b28} showed that eye gaze, when combined with speech as part of a feature fusion, single support vector regression system, could improve arousal prediction compared to that of unimodal speech (3.5\% relative performance improvement), while model fusion improved valence prediction compared to unimodal speech (19.5\% relative performance improvement). Psychopathological affective computing work incorporating eye-based features as part of multimodal approaches include post traumatic stress disorder estimation \cite{b29} and depression recognition \cite{b30, b31}.

The affective computing community is acknowledging the potential for eye-based cues for system development. However, the full benefit of eye-based cues estimated from video for affective computing purposes is not yet known, despite audio-visual data and computer vision tools that are now widely accessible for feature extraction.

\section{Data and Tools}
The RECOLA corpus \cite{b21} is used as the experimental data set for this work. RECOLA is an affective data set comprised of audio-visual and physiological recordings of subjects cooperating on a task and communicating in French. Arousal and valence annotations, ranging from -1.0 to +1.0, are provided with the set in discrete-continuous-time at a rate of 25 values per second. Subject meta-data, such as mother tongue and sex, are also provided. Recordings of 23 subjects available in the set were paritioned into training, validation and test sets with the aim of matching the distributions used in \cite{b34}. Specifically, the training set is comprised of subjects [P16, P17, P19, P21, P23, P26, P30, P65], the validation set includes subjects [P25, P28, P34, P37, P41, P48, P56, P58], and the test set includes subjects [P39, P42, P43, P45, P46, P62, P64].

A Dell Precision Tower 3620 with Intel Core i7 6700K, 8 GB RAM and NVIDIA Quadro M2000 Graphics Card, Ubuntu 16.04 computing system was used for the experiments. Key software for the experiments included: OpenFace (version 2.0.6) \cite{b20} for gaze, eye blink/closure and pupils estimation from video, the CUDA RecurREnt Neural Network Toolkit (version 0.2 rc1) \cite{b32} for BLSTM-RNN model training, and the R programming language/interpreter (version 3.4.4) \cite{b33} for the proposed feature set extraction and statistical analyses. Additional software packages and required software dependencies are detailed in the repository accompanying this work.

\section{Experiment Design}

\subsection{Initial Eye-based Features}
The initial eye-based features were extracted from 6 binary [direct gaze, gaze approach, eyes fixated, eye closure/blink, pupil dilation and pupil constriction] and 8 numerical [eye blink intensity,  pupil diameter, $\Delta$pupil diameter, $x$ and $y$ gaze angles, $\Delta x$ and $\Delta y$ gaze angles, and pupil diameter] low level descriptors (LLDs). The LLDs were gathered frame-wise from each subject video recording. The pupil estimation was based on the left eye, due to OpenFace's model implementation, and gaze angles were measured in radians. All but eye closure/blink, blink intensity and $x$ and $y$ gaze angles LLD features required calculation in this work. The direct gaze binary variable required a human coder to view cropped images of individual's faces and provide true/false ratings based on whether they thought the interacting subject was looking at the interlocutor (direct gazing) or not.

For the pupil modality, mid-level features were gathered using an 8 second time window (200 frames at 25 frames per second) moved forward at a rate of 1 frame per interval. This particular time window and rate are commonly employed for continuous affect prediction \cite{b35}. Specifically, 10-order Daubechies \cite{b36} discrete wavelet transform features are gathered at 7 levels of decomposition, the maximum possible decomposition for the time window used, based on \cite{b10}.

Following the low- and mid-level descriptor feature extraction, statistics were applied to achieve an initial set of 292 features that comprised of 69 eye gaze features, 209 pupillometry features and 14 eye closure features. The initial feature list included:
\newline
\newline
$\bullet$ Direct gaze, pupil dilation and pupil constriction (12 features):
\newline
ratio, time seconds: mean, max, total
\newline
\newline
$\bullet$ Gaze approach, eyes fixated and eye closure/blink (14 features):
\newline
ratio, time seconds: min (min not applied for gaze approach), median, mean, max
\newline
\newline
$\bullet$ Pupil diameter, $\Delta$pupil diameter, $x$, $y$ gaze angles, $\Delta x$ and $\Delta y$ gaze angles (84 features):
\newline
min, max, mean, median, quartile 1, quartile 3, skewness, kurtosis, standard deviation, inter-quartile range (IQR) 1-2, IQR 2-3, IQR 1-3, linear regression slope, linear regression intercept
\newline
\newline
$\bullet$ Eye blink intensity (9 features):
\newline
 max, mean, median, quartile 3, standard deviation, IQR 1-2, IQR 2-3, linear regression slope, linear regression intercept
\newline
\newline
$\bullet$ 10-order Daubechies scale and approximation wavelet coefficients at 7 levels of decomposition (173 features):
\newline
min, max, median, quartile 1, quartile 3, skewness, kurtosis (kurtosis not applied for final scale and approximation wavelet coefficients), standard deviation, IQR 1-2, IQR 2-3, IQR 1-3, RMS, zero crossing rate (ZCR) (ZCR not applied to scale coefficients)

\subsection{BLSTM-RNN Training and Evaluation Method}
In this work, BLSTM-RNN was used to train models for feature set appraisal. The training method used largely follows that of Ringeval et al. \cite{b34}. Single-task models were trained using BLSTM-RNN with 2 hidden layers, each with 40 and 30 nodes respectively, with a sum-of-squared-errors (SSE) objective function. All input features and regression targets were standardised using the parameters mean and standard deviation, computed on the training set. The network learning rates were set at $10^{-5}$ and a random seed of 1787452436 was used throughout the experiments. Gaussian noise with a standard deviation 0.1 was added to all input features prior to training. BLSTM-RNN models were trained for a maximum of 100 epochs, however, training was stopped when no performance increase (lower SSE) was observed on the validation set after 10 epochs.

Following the training phase, network models were evaluated and selected using concordance correlation coefficient (CCC) \cite{b41, b42}, where higher CCC is better. The CCC measure penalises correlated time-series by applying a penalty of mean-squared error as in \eqref{CCC-def}, where \textit{x} represents predicted values, \textit{y} represents ground-truth values, $\sigma_{xy}$ is the covariance, $\sigma^2$ is the variance and $\mu$ is the mean.

\begin{equation}
  \label{CCC-def}
  CCC = \frac{2 \sigma_{xy}}{\sigma^2_{x} + \sigma^2_{y} + (\mu_{x} - \mu_{y})^2} \
\end{equation}

\subsection{Feature Selection}
The feature selection approach that was taken followed a simple approach of mutual information (MI) estimation to regression target-based filtering. MI is ``the amount of information that one random variable contains about another random variable'' [39, p.18]. An additional component that was considered for the feature selection process included a now common affect learning parameter, ground-truth backward time-shift. It is accepted in the literature that human annotators produce a delay when providing their ground-truth ratings. To mitigate for this, researchers now incorporate time-shifting of ground-truth (or gold-standard) values back in time for continuous affect prediction purposes \cite{b31,b35,b40,b41,b42}. However, ground-truth time-shifts are evaluated on the validation set using model performance CCC and the ground-truth time-shifting may occur prior to achieving the best feature set, and therefore, the best inputs to models for ground-truth time-shift assessment may not be present. When carrying out feature selection in this work, MI was estimated between the input features and ground-truth regression targets \textit{before}, \textit{during}, and \textit{after} ground-truth time-shifting has occurred. 

The MI thresholds evaluated were 0.1, 0.15, 0.2. A total of  23 ground-truth time-shifts were applied ranging from 0 (not applied) to 4.4 seconds, altered in backward shifts of 0.2 seconds. Features that have a MI less than the threshold under test were removed as these features were deemed independent of arousal or valence, and therefore poor predictors. For the ground-truth time-shift iterated MI selection, the threshold that provided the best performing feature set \textit{before} any ground-truth time-shift was applied was again used \textit{during} feature selection after each backward time-shift iteration. Ground-truth time-shift is referred to as $D_{s}$ for the remainder of this work, where $s$ is the value in seconds for the delay $D$ applied to ground-truth.

\subsection{Practical Significance Evaluation}
In order to assess the practical significance of the eye-based feature sets following feature selection, two steps are taken. Firstly, group-of-humans-level performance estimates were calculated using the RECOLA training set arousal and valence annotations. These estimates consist of averages for annotator-to-annotator CCC for the training set arousal and valence rating provided. The estimates are important for continuous arousal and valence prediction in order to have a practical baseline for automatic systems. The results of these calculations are group-of-humans-level arousal prediction CCC = 0.341, and valence prediction CCC = 0.383.

While comparisons between the group-of-humans-level performance and that of the eye-based cues validation set results can be made, this is an unfair comparision. The human coders have access to additional modalities to reach their decision while the eye-based models and features have been tuned on the validation set. It is clear that advantages from both comparison groups have been taken away, for example, both human generalisation and computing machine capabilities are not being taken advantage of. Therefore, the second step in the practical significance evaluation is to offer as much of a like-for-like comparison as possible between the group-of-humans-level and eye-based performances. Since the human annotators have the advantage of both audio and visual data for decoding and rating affect, it was decided that speech would be combined with the eye-based features for practical evaluation and comparison to that of the human coders. The eGeMAPS \cite{b43} speech feature set was used in the experiments and it was combined with the eye-based features using early feature fusion. Performance of bimodal systems were assessed versus unimodal speech on the validation set, and, if improvements were observed, a test set pass was carried out for the speech and eye-based systems, followed by group-of-humans-level performance comparisons.

\subsection{Eye-based Arousal and Valence Feature Set Proposals}
The experiments culminate with the results for eye-based affective computing sets, intended for the continuous prediction of arousal and valence affect dimensions based on video input. The final ground-truth time-shift parameter required for use with the sets, along with the proportions of retained features, are given along with discussion of the final feature sets. The top 20 performers from each of the arousal and valence sets, ranked in terms of both linear, absolute value PCC, and nonlinear, MI, metrics, are provided for the final, proposed feature sets. The full list of features will be made available in the repository for this work. 

\section{Results and Discussion}

\subsection{Feature Selection}
The results are given in Table \ref{table1} for feature selection carried out \textit{before} application of $D_{s}$. Table \ref{table1} shows that this feature selection technique is effective. Performance increases in terms of validation set CCC along with feature set size reductions are always observed compared to when feature selection was not applied. These results mean that a more optimal sub-set of features can be used for further feature selection incorporating $D_{s}$. The sets gathered from this experiment include \textit{before} condition feature sets for arousal and valence of sizes 147 and 152 respectively. Additionally, parameters resulting from this experiment include MI threshold values, 0.15 for arousal and 0.2 for valence, which are used for the $D_{s}$ iterated feature selection from the feature sets at each iteration.

\begin{table}[htbp]
\caption{Validation Set Feature Selection BLSTM-RNN Results Before Ground-truth Time-shift}
\begin{center}
\begin{tabular}{|c|c|c|c|c|c|c|}
\hline
\multirow{2}{*}{\textbf{\begin{tabular}{c}MI $<$\\Filter\end{tabular}}}&\multicolumn{3}{|c|}{\textbf{Arousal}}&\multicolumn{3}{|c|}{\textbf{Valence}}\\\cline{2-7} & \textbf{SSE} & \textbf{CCC} & \textbf{Features} & \textbf{SSE} & \textbf{CCC} & \textbf{Features} \\
\hline
N/A  & 0.368 & 0.106 & 292 & 0.417 & 0.000 & 292 \\
0.1  & 0.361 & 0.15  & 182 & 0.414 & 0.029 & 198 \\
0.15 & 0.346 & \textbf{0.188} & 147 & 0.414 & 0.032 & 152 \\
0.2  & 0.352 & 0.187 & 99  & 0.405 & \textbf{0.058} & 124 \\
\hline
\end{tabular}
\label{table1}
\end{center}
\end{table}

\begin{figure*}[htbp]
\centering
\subfloat[]{\includegraphics[width=2.5in]{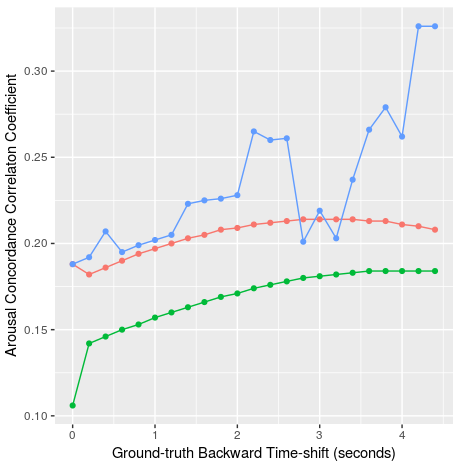}
\label{fig_a}}
\hfill
\subfloat[]{\includegraphics[width=4in, height=2.5in]{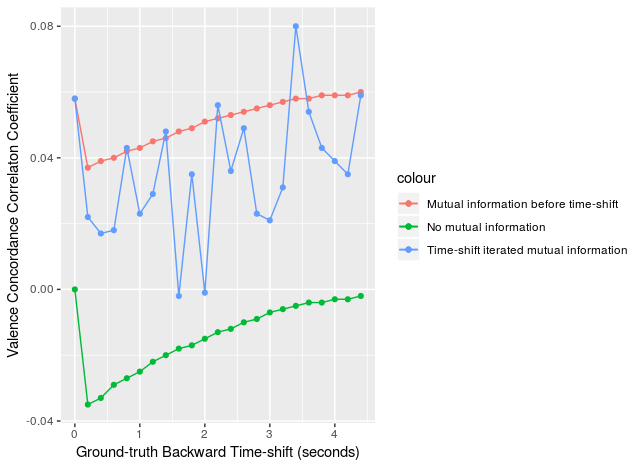}
\label{fig_b}}
\caption{Arousal (a) and valence (b) validation set CCC scores under different ground-truth backward time-shift ($D_{s}$) conditions. $D_{s}$ shifts evaluated ranged from 0 (not applied) to 4.4 seconds, in increments of 0.2 seconds.}
\label{figure1}
\end{figure*}

The results depicted in Fig.~\ref{figure1} (a) showed that the $D_{s}$ iterated MI feature selection (blue line) for arousal performs better than the other methods at this experimental stage. Additionally, assessing how much greater the iterated MI feature selection was compared to the MI selection (red line) \textit{before} $D_{s}$ shifting was found to be statistically significant (Wilcoxon rank sum test, W = 378.5, p-value = 0.006). The highest performing $D_{s}$ value for arousal was 4.4 seconds, with an eye-based feature count of 151 features that was achieved using the MI threshold = 0.15. The validation set arousal CCC for this system was 0.326. Another result from this experiment included the highest performing $D_{s}$ for the group where no MI feature selection was applied, which was 4.4 seconds. Therefore, where MI will be applied \textit{after} $D_{s}$, a 4.4 seconds $D_{s}$ is applied to the arousal ratings first.

Fig.~\ref{figure1} (b) shows the $D_{s}$ iterated MI feature selection for valence. It is clear from this graph that not utilising feature selection prior to, or during $D_{s}$ shifting, model building and evaluation, can be detrimental. Both MI feature selection \textit{before} and in the iterated fashion, performed better than when MI was not applied and $D_{s}$ effects were evaluated. The top performer in terms of CCC came from the $D_{s}$ iterated MI group, a value of 0.08, and this was achieved using $D_{s}$ = 3.4 seconds applied to valence ground-truth. The top performing feature set size includes 128 of the originally proposed 292 features, gathered using an MI threshold of 0.2. Due to the difficultly in concluding which $D_{s}$ performed best for the group where no MI feature selection was applied, it was decided against applying any feature selection \textit{after} $D_{s}$ for this group. 

The final MI feature selection to be applied \textit{after} $D_{s}$ shifting could only be applied to the arousal set. The results for this feature selection are given in Table \ref{table2}. The poor performance of the valence feature set without MI applied prior to $D_{s}$ shifting (performance always dropped) resulted in no plausible selection for $D_{s}$ and MI selection \textit{after} shifting for this affect dimension. The top performing result in Table \ref{table2} is the same as from the $D_{s}$ iterated MI selection, with matching feature set size, SSE, and CCC. These early results indicate that, for arousal, applying feature selection \textit{during}, or \textit{after} $D_{s}$ shifting makes no difference in terms of CCC performance. An interesting point to note from Table \ref{table2} is that the the top performing system, achieving a CCC of 0.326, fell just 4.4\% short of the group-of-humans-level CCC performance estimate of 0.341, which provides early evidence of promise for this eye-based feature set for arousal prediction.

\begin{table}[htbp]
\caption{Validation Set Feature Selection BLSTM-RNN Results After Ground-truth Time-shift Applied}
\begin{center}
\begin{tabular}{|c|c|c|c|}
\hline
\multirow{2}{*}{\textbf{\begin{tabular}{c}MI $<$\\Filter\end{tabular}}}&\multicolumn{3}{|c|}{\textbf{Arousal}}\\\cline{2-4} & \textbf{SSE} & \textbf{CCC} & \textbf{Features} \\
\hline
N/A  & 0.347 & 0.184 & 292 \\
0.1  & 0.327 & 0.277 & 187 \\
0.15 & 0.313 & \textbf{0.326} & 151 \\
0.2  & 0.334 & 0.205 & 111 \\
\hline
\end{tabular}
\label{table2}
\end{center}
\end{table}

\subsection{Practical Significance Evaluation}
The results for the practical significance evaluation are given in Table \ref{table3} for the speech feature set alone and when combined with the final eye-based feature sets. The same $D_{s}$ shift and MI feature selection values as used for the eye-based features were applied. The results show that when eye-based features are combined with speech, performance benefits can be achieved. A relative increase of 9.19\% above that of uimodal speech is observed in Table \ref{table3}. Due to the increase in performance of the bimodal system for arousal, a test set pass was performed for this affect dimension. Unfortunately, the bimodal valence system did not perform well, there was a performance decrease found for this dimension. Further experimental work is required for valence prediction using eye-based cues from video, based on the results presented they appear unsuitable for valence prediction either in the presence or absence of speech. The test set result CCC of 0.72 for arousal achieved by the speech and eye-based system in Table \ref{table3} compares favourably to that of the group-of-humans-level CCC of 0.341, a 111\% performance increase relative to the group of human annotators. This out-of-sample result indicates promise for eye-based cues for continuous arousal prediction, especially when considered in multimodal systems. 

\begin{table}[htbp]
\caption{Final BLSTM-RNN Results For Systems Including Speech}
\begin{center}
\begin{tabular}{|c|c|c|c|c|}
\hline
\multirow{2}{*}{\textbf{\begin{tabular}{c}System\\(Evaluation)\end{tabular}}}&\multicolumn{2}{|c|}{\textbf{Arousal}}&\multicolumn{2}{|c|}{\textbf{Valence}}\\\cline{2-5} & \textbf{SSE} & \textbf{CCC} & \textbf{SSE} & \textbf{CCC} \\
\hline
\begin{tabular}{c}Speech-based\\(Validation)\end{tabular}& 0.192 & 0.675 & 0.391 & \textbf{0.103} \\
\begin{tabular}{c}Speech \& Eye-based\\(Validation)\end{tabular}& 0.17 & \textbf{0.737} & 0.402 &  0.059 \\
\begin{tabular}{c}Speech \& Eye-based\\(Test)\end{tabular}& - & 0.72 & - & - \\
\hline
\end{tabular}
\label{table3}
\end{center}
\end{table}

\subsection{Eye-based Arousal and Valence Feature Set Proposals}
The final 151-dimensional eye-based arousal feature set produced from this work was gathered using the MI threshold set to 0.15 during $D_{s}$ iterated feature selection, with $D_{s}$ = 4.4 seconds. The retention proportion from each of the information chananels for the final set are as follows: eye gaze (55 of 69 features), pupillometry (85 of 209 features) and eye closure/blink (11 of 14 features). The retention proportions for the 128-dimensional eye-based valence feature set included: eye gaze (46 of 69 features), pupillometry (73 of 209 features) and eye closure/blink (9 of 14 features). This feature set was gathered using $D_{s}$ = 3.4 seconds along with the MI threshold set to 0.2 which was applied during $D_{s}$ iterated feature selection. For both arousal and valence feature sets, it is interesting to note that all 4 direct gaze binary-based features were retained in the final sets, and all of both pupil constriction and pupil dilation binary-based features were removed from the final sets using the MI feature selection.

\begin{table*}[htbp]
\caption{Top Features Ranked by Correlation and Mutual Information With Arousal and Valence}
\begin{center}
\begin{tabular}{|c|c|c|c|c|c|c|c|}
\hline
\multicolumn{4}{|c|}{\textbf{Arousal}}&\multicolumn{4}{|c|}{\textbf{Valence}}\\
\hline
\textbf{Feature} & \textbf{PCC} & \textbf{Feature} & \textbf{MI} & \textbf{Feature} & \textbf{PCC} & \textbf{Feature} & \textbf{MI} \\
\hline
 gaze x max & 0.361 & gaze x max & 0.57 & scale coeffs l1 max & 0.321 & $\Delta$ gaze y max & 0.594 \\
 gaze x quartile 3 & 0.33 & $\Delta$ gaze y max & 0.547 & scale coeffs l2 max & 0.319 & $\Delta$ gaze y min& 0.568 \\
 gaze x mean & 0.304 & $\Delta$ gaze y min & 0.539 &\begin{tabular}{c}pupil diameter\\mm max\end{tabular} & 0.315 & gaze x max & 0.548 \\
 gaze x median & 0.299 & gaze y min & 0.514 & gaze x max & 0.306 & $\Delta$ gaze x max& 0.537 \\
 gaze y min & -0.266 &\begin{tabular}{c}eye blink intensity\\max\end{tabular}& 0.508 & scale coeffs l3 max & 0.295 & $\Delta$ gaze x min & 0.528 \\
 \begin{tabular}{c}$\Delta$ gaze y inter-quartile\\range (IQR) 1-3\end{tabular} & 0.242 & $\Delta$ gaze x min & 0.506 & gaze x quartile 3 & 0.287 & gaze x min & 0.498 \\
 $\Delta$ gaze y quartile 3 & 0.236 & $\Delta$ gaze x max & 0.505 & gaze x median & 0.281 & gaze y max & 0.498 \\
gaze x quartile 1 & 0.236 & gaze x min & 0.496 & gaze x mean & 0.278 & \begin{tabular}{c}eye blink intensity\\max\end{tabular} & 0.495 \\
 $\Delta$ gaze y IQR 1-2 & 0.234 & gaze y max & 0.469 & scale coeffs l4 max & 0.273 & gaze y min & 0.493 \\
 $\Delta$ gaze y IQR 2-3 & 0.231 &\begin{tabular}{c}$\Delta$ pupil diameter\\mm max\end{tabular}& 0.421 & wavelet coeffs l2 SD & 0.267 & \begin{tabular}{c}$\Delta$ pupil diameter\\mm min\end{tabular} & 0.482 \\
 $\Delta$ gaze y quartile 1 & -0.226 & gaze y median & 0.41 &\begin{tabular}{c}wavelet coeffs l2\\RMS\end{tabular}& 0.267 & \begin{tabular}{c}$\Delta$ pupil diameter\\mm max\end{tabular} & 0.437 \\
 \begin{tabular}{c}gaze x standard\\deviation (SD)\end{tabular} & 0.225 & gaze x quartile 3 & 0.404 & \begin{tabular}{c}pupil diameter mm\\quartile 3\end{tabular} & 0.249 & \begin{tabular}{c}pupil diameter\\mm max\end{tabular} & 0.414 \\
 direct gaze time ratio & 0.224 &\begin{tabular}{c}$\Delta$ pupil diameter\\mm min\end{tabular}& 0.402 & scale coeffs l5 max & 0.24 & gaze y quartile 3 & 0.407 \\
\begin{tabular}{c}wavelet coeffs l3\\RMS\end{tabular}& 0.222 & gaze y quartile 3 & 0.4 &\begin{tabular}{c}scale coeffs l1\\quartile 3\end{tabular}& 0.236 & gaze y median & 0.403 \\
 wavelet coeffs l3 SD  & 0.222 & $\Delta$ gaze y SD & 0.4 & gaze x quartile 1 & 0.237 & max gaze fixation time & 0.399 \\
 \begin{tabular}{c}pupil diameter\\mm max\end{tabular} & 0.21 & gaze x median & 0.389  & gaze y min & -0.226 & gaze x median & 0.397 \\
 gaze y quartile 1 & -0.207 & gaze x quartile 1 & 0.388 &\begin{tabular}{c}scale coeffs l2\\quartile 3\end{tabular}& 0.226 & $\Delta$ gaze y SD & 0.396 \\
 gaze y mean & -0.203 & gaze y quartile 1 & 0.385  & wavelet coeffs l2 max & 0.224 & gaze x quartile 3 & 0.395 \\
 gaze y SD & 0.2  & max gaze fixation time & 0.373 & $\Delta$ gaze y IQR 1-2 & 0.218 & scale coeffs l1 max & 0.392 \\
wavelet coeffs l2 SD & 0.195 & max eyes closed time & 0.36 & scale coeffs l2 max & 0.218 & gaze y quartile 1 & 0.39 \\
\hline
\end{tabular}
\label{table4}
\end{center}
\end{table*}

The top 20 performers for the final feature sets are given in Table \ref{table4}. The top-ranked performer of the arousal features is \textit{gaze x max}, which is highest both in terms of PCC (0.361) and MI (0.57). The top performing valence features include Daubechies \textit{scale coefficients l1 max}, which achieved a PCC of 0.321, and $\Delta$\textit{gaze y max}, which achieved a MI of 0.594. The top performers for arousal contain eye gaze features for 16 of the top 20 in terms of PCC and 17 of the top 20 in terms of MI. For the valence dimension, 13 of the top 20 features ranked by PCC are provided by pupil-based features, with the majority of these features coming from Daubechies wavelet coefficients (11 of 13). In terms of MI with valence, 15 of the top 20 features shown in Table \ref{table4} are provided by eye gaze features. 

\section{Conclusions and Future Work}
This work investigated eye-based cues, gathered from video, intended for use in affective computing. Cues from gaze, pupillometry and eye closure were proposed and feature vectors were refined for continuous arousal and valence prediction using ground-truth backward time-shifting, feature selection and evaluation using BLSTM-RNN. Performance comparable to that of group-of-humans-level arousal CCC was achieved on the validation set for the eye-based cues on their own. Additionally, the results obtained show the benefit of combining the eye-based cues with speech for arousal prediction; the CCC for the bimodal system was 0.72 compared to the group-of-humans baseline of 0.341 on the RECOLA test set. Eye gaze features were shown to be particularly salient for arousal prediction from the eye-based cues with the majority of top 20 performers as measured by both linear and nonlinear relationships with arousal provided from gaze. The validation set performance of the eye-based features for valence was poor when combined with speech, providing performance degradation in terms of CCC compared to unimodal speech. This study shows that valence features from eye-based cues gathered from video require further investigation prior to practical application. Potetential avenues for further investigation of eye-based features for valence may include more advanced feature selection, feature fusion, automatic feature learning and different regression techniques for prediction. The majority of top performers for the final valence feature set comprised of Daubechies wavelet features for linear relationships with valence and eye gaze features for nonlinear relationships with valence.

Some limitations of this study include the nonoptimal fusion and ground-truth time-shift of the speech and eye-based cues, the human coder required to provide direct gaze binary annotations and the lack of consideration for eye-based cues of ground-truth providing annotators. Future work includes incorporating the proposed eye-based cues for arousal into multimodal systems that include speech, facial expression and head pose, as well as addressing the optimal fusion strategy and regression technique for these cues. The development of a deep learning-based direct gaze estimation model from video and a pilot study of ground-truth provided with recordings of annotator eye measurements are also planned.

\vspace{12pt}
\color{red}

\end{document}